# Nanoscale thermal conductivity of Kapton-derived carbonaceous materials

K. Kondratenko*, D. Hourlier, D. Vuillaume, S. Lenfant *

*Institut d'Electronique de Microélectronique et de Nanotechnologie (IEMN), CNRS, Avenue Poincaré, 59652 Villeneuve d'Ascq, France.*

* Corresponding authors: kirill.kondratenko@iemn.fr, stephane.lenfant@iemn.fr

___________________________________________________________________________

This study exploits the nanoscale resolution of Scanning Thermal Microscopy (SThM) to reveal inhomogeneous nature of thermal properties of carbon-derived materials issued from thermal conversion of the most commonly known polyimide (PI), Kapton®. This information is otherwise inaccessible if conventional thermal characterization techniques are used due to their limited spatial resolution. Kapton films were pyrolyzed in an inert atmosphere to produce carbon-based residues with varying degree of conversion to free $sp^2$ disordered carbon. The thermal conductivity of carbon materials ranges from 0.2 to 2 $Wm^{-1}K^{-1}$, depending on the temperature of the carbonization process (varied between 500°C and 1200°C). For quantitative measurements of thermal conductivity, the Null Point SThM (NP-SThM) technique is used in order to avoid unwanted effects as the parasitic heat flows through the air and the probe cantilever. It was found that NP SThM data for bulk materials are in excellent agreement with results obtained through more traditional techniques, namely photo-thermal radiometry, flash laser analysis and micro-Raman thermometry. This allowed us to use the NP-SThM technique to differentiate structural heterogeneities or imperfections at the surface of the pyrolysed Kapton on the basis of measured local thermal conductivity.

___________________________________________________________________________

# 1. Introduction

With increasing degree of integration and downscaling of components, thermal management becomes one of the key parameters of improving performance and reliability of electronic devices, especially with the advent of 3D integrated circuits[1]. Carbon-based materials have demonstrated promising properties of heat transport with a thermal conductivity surpassing 3000 $Wm^{-1}K^{-1}$ for some allotropic variants[2]. Carbon nanotubes[3] and graphene[4] are proposed as high performance materials for thermal vias and heat spreaders. Polyimide films (commercialized as Kapton®) have aroused interest as material for high-tech industry, owing to their mechanical properties, high thermal stability, favorable dielectric properties, chemical inertness and compatibility with semiconducting materials. Today, Kapton film continues to draw attention as a precursor for materials consisting of disordered $sp^2$ carbon obtained upon treatment at 600 °C and higher, which further forms graphitic-like material upon even higher temperature (greater than 1000°C[5]). This treatment, if performed in an inert atmosphere, results in thermal decomposition of polyimide accompanied by structural transformation[6], shrinkage and mass reduction due to the elimination of heteroatoms. The resulting product is a residue mostly consisting of carbon[7]. These materials are potential candidates for terahertz absorbers[8] used as detectors and shielding materials, as well as for heat dissipating applications.

It has been shown that the thermal conductivity can be changed in a wide range of almost 4 orders of magnitude by various heat treatments, from 0.2 $Wm^{-1}K^{-1}$ for untreated Kapton up to 1700 $Wm^{-1}K^{-1}$ for samples with high degree of graphitization[9]. This makes carbonized Kapton a convenient reference material for several thermal characterization techniques, which was recently exploited by our group in a study of Kapton-derived carbonaceous materials with three different techniques[10]: laser flash analysis, photothermal radiometry and Raman thermometry.

Scanning Thermal Microscopy (SThM) is a known method for the characterization of local thermal conductivity of organic thin films[11–13], nanoobjects[14], 2D materials[15], 1D materials[16], monolayers[17] and down to single molecules[18–20]. In this technique, the thin film resistor in the tip of scanning probe microscope is essentially utilized as a thermometer that allows obtaining decananometric lateral resolution[21]. In active mode SThM, the resistive probe is heated by DC bias, which allows not only to measure temperature (which may be useful in investigating samples with integrated heat source) but also to induce heat flow from the tip into the sample under study. This can then be used to find the thermal conductance of materials.

The SThM requires little or no sample preparation in contrast to other traditional techniques, such as 3ω Joule heating which requires the deposition of electrodes[22], sometimes of unconventional shape[23]. Other techniques which utilize optical excitation require additional steps as well: laser flash analysis requires the deposition of a graphite layer to absorb the laser light and to emit the infrared radiation for detection[24]. In time-domain thermoreflectance measurements (TDTR), a thin metal film is usually deposited for similar reasons[25,26]. Even though these techniques provide accurate results for bulk materials, submicron lateral resolution cannot be achieved.

In order to obtain reliable quantitative information on thermal conductivity, the measurement of the heat flow from the SThM tip to the sample is of considerable importance[27]. This measurement is complicated due to the parasitic heat flow through air and the SThM probe cantilever, which are difficult to estimate experimentally. Kim *et al.* have developed a calibration procedure (Null Point SThM)[28] for the SThM measurements by using reference materials of known thermal conductivity. In this study, we used the NP SThM technique to probe nanoscale inhomogeneities of the thermal conductivity of Kapton-derived materials.

## 2. Experimental

We prepared our samples from commercially available yellow-orange Kapton<sup>®</sup> HN polyimide films of thickness 125 µm, (Du Pont de Nemours, France) following a procedure previously described elsewhere[8]. Carbonization of Kapton results in a slight decrease of the film thickness (reaching 117 µm for the sample treated at 1200°C). The thermal characterization of carbonized Kapton was performed on a Bruker Dimension ICON with the Anasys SThM application module in DC mode. Measurements were performed in air-conditioned environment at T=22.5°C and relative humidity of 35-40%. Kelvin NanoTechnology (KNT) SThM probes were used, the heating element consists of a thin Pd film resistor inserted in a Wheatstone bridge circuit with DC bias varied between 0.7 and 1.2 V.

Thermal conductivity of Kapton-derived carbonaceous materials was obtained by Null Point (NP) SThM[28], which allows removing of the parasitic contribution of heat conduction through the environment. Before contact, temperature of the tip of SThM probe is influenced by heat losses through air and probe cantilever. Upon physical contact with the sample, the heat flow through the tip-sample interface is added to the total tip heat dissipation. By continuously recording the tip temperature before and during heat exchange with the sample and in conjunction with a careful calibration procedure, this technique provides a quantitative estimate of the thermal conductivity.

## 3. Results

### 3.1 Thermal conductivity by NP SThM

In this protocol, the heated SThM tip (temperature of the tip is controlled by the DC bias of the Wheatstone bridge) is brought in contact with the surface of the sample by performing a ramp in Z direction of the scanning probe microscope scanner. At the moment of contact, a rapid temperature change occurs due to induced heat flow from the heated tip into the sample. This temperature change is recorded as a voltage differential at the output of the Wheatstone bridge of the SThM module, which allows us to obtain a sequence of temperature differential values for different probe temperatures.

We calculate the thermal conductivity κ by using the following relation[28]:

$$T_C - T_{amb} = \left[\alpha \frac{1}{\kappa} + \beta\right](T_{NC} - T_C) \qquad (1)$$

where $T_C$ is the probe tip temperature at contact, i.e. when the tip is in thermal equilibrium with the sample surface (visible for Z-$Z_C$>0 in the Figure 1 (a)), $T_{NC}$ is the tip temperature just before contact with the sample surface, $T_{amb}$ is the ambient temperature, the calibration coefficients α and β are related to thermal contact area, i.e. tip and sample geometry as well as tip-sample thermal conductance[28].

The values of $T_C$-$T_{amb}$ are plotted as a function of $T_{NC}$-$T_C$ and we perform a linear fit of this plot to extract its slope, χ, which is inversely proportional to the thermal conductivity. The thermal conductivity is obtained through prior calibration of the system using reference materials. For this purpose, reference materials are Si wafers (we assume its conductivity to 148 Wm$^{-1}$K$^{-1}$, with the native silicon oxide contribution being negligible[28]) and borosilicate microscope glass slide (κ = 1.3 Wm$^{-1}$K$^{-1}$), see supporting information (Figure S1).

Since the quantitative estimation of thermal conductivity is extremely sensitive to the calibration parameters, we should consider one of the main sources of error in this type of measurement, which is the variation of surface area of thermal contact between the heated tip and the sample[29]. In order to decrease its contribution, one should pay attention to the surface roughness of the sample. In our case, for heat-treated Kapton materials, the root mean square (RMS) surface roughness is in the order of 1 nm (Figure S2 in the SI). In an attempt to improve the measurement accuracy, we used a 5 by 5 grid pattern with a pitch of 100 nm. Figure 1 (a) shows an example of data collected for Kapton derived samples heated at 1200°C and 500°C at 1.1V Wheatstone bridge bias. Figure 1(b) shows the plots of $T_C$-$T_{amb}$ vs. $T_{NC}$-$T_C$ corresponding to one grid (25 measurements per Wheatstone bridge bias value) for the same samples obtained during one measurement session. The data for the other temperatures are given in the Figure S3 in the supporting information.

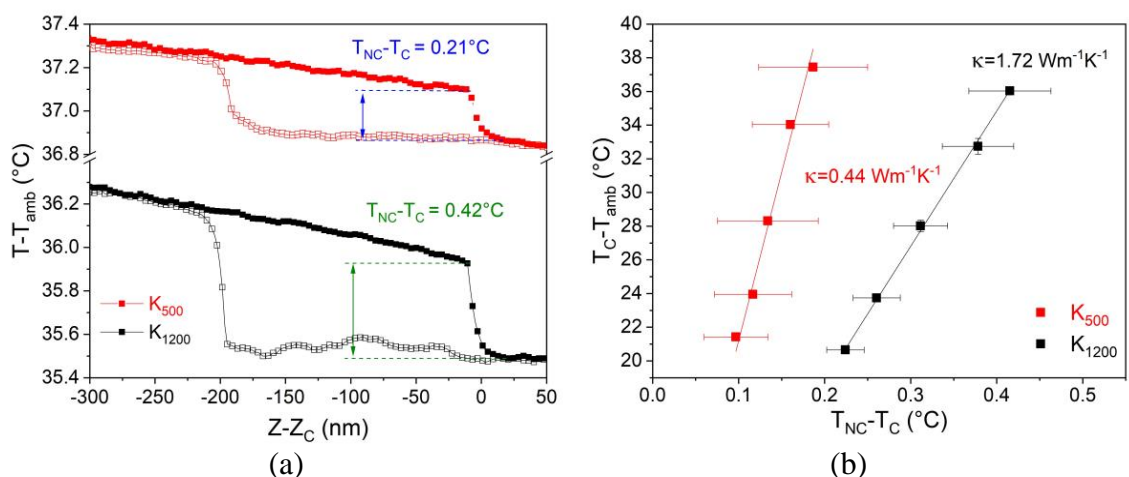

**Figure 1**: (a) Example of NP SThM temperature measurement data (tip temperature as a function of scanner displacement in Z direction): $Z_C$ is scanner coordinate at contact. Wheatstone bridge voltage is 1.1V. Solid symbols correspond to extend of the cantilever, open - to retract. (b) Plot of $T_C$-$T_{amb}$ as a function of $T_{NC}$-$T_C$ for samples of Kapton heated at 1200°C (black) and 500°C (red), Wheatstone bridge bias was varied between 0.7V and 1.1V in steps of 0.1V. Thermal conductivity marked on the Figure and calculated from Eq. 1 using calibration parameters: α=69.93 Wm$^{-1}$K$^{-1}$ and β=39.56 K/K (see Figure S2 in the SI).

## 3.2 Local thermal conductivity

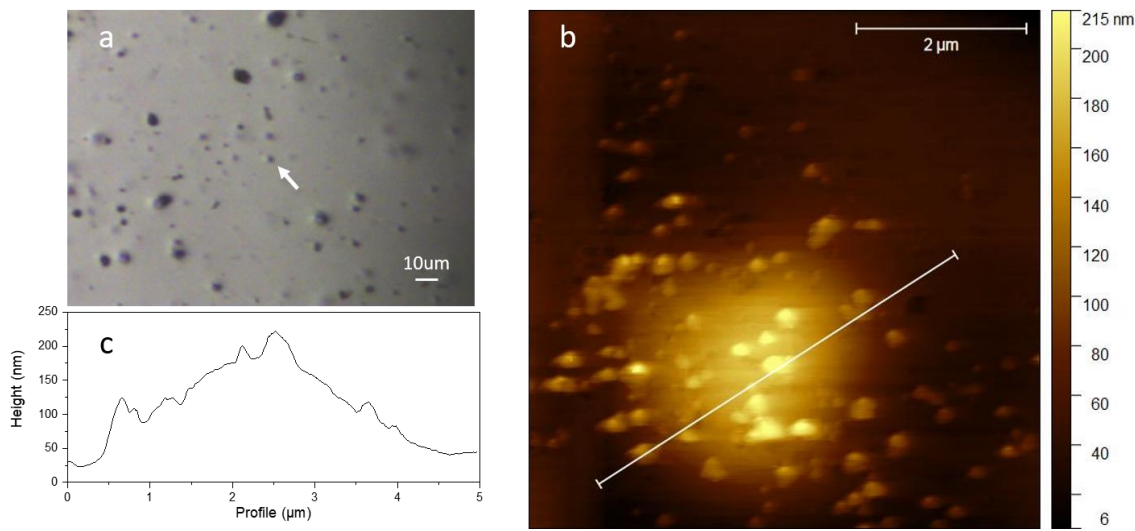

**Figure 2**: Topological features of sample treated at 1200°C: optical image (a); topography recorded by SThM tip in contact mode (b); profile extracted from topographical image (c).

The optical micrograph presented in Figure 3(a) shows the heat treatment induced changes in the Kapton film and reveals the presence of nodules taller than 100 nm Figure 3 (b) as shown by the topographic image recorded simultaneously during the SThM scan.

For the local thermal conductivity of the 1200°C-Kapton-derived materials, measurements were conducted using a high lateral resolution of NP SThM (ca. 100 nm). The topology in contact mode and the thermal voltage are simultaneously measured and recorded (Figures 2 (b) and 3, respectively). The nodule is surrounded by grainy features visible in the topological image. The thermal image (Figure 3) reveals noticeable difference between the nodule and the surrounding area, which prompted us to perform local thermal conductivity measurements. This thermal image alone however provides only qualitative information of the thermal conductance of the sample and is not suitable for calculation of the local thermal conductivity due to heat losses and thermal contact area fluctuation during the surface scan.

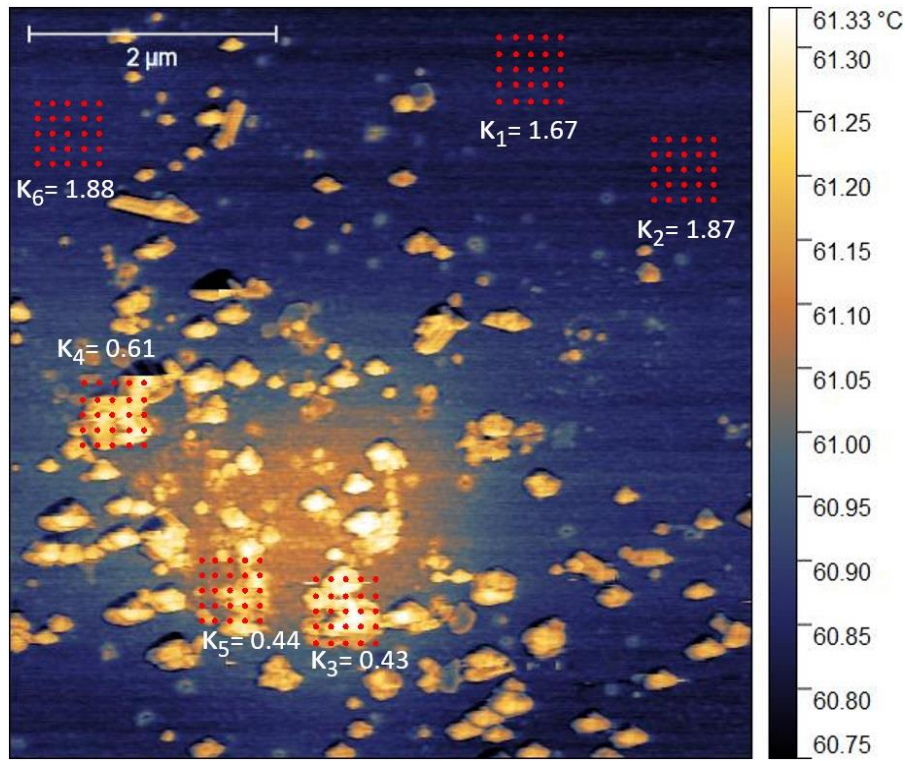

**Figure 3**: Thermal image (tip temperature) of the topographic feature presented in Figure 2. Thermal conductivity values are in $Wm^{-1}K^{-1}$. Wheatstone bridge voltage is 1.2V.

NP SThM was performed on selected spots corresponding to the nodule and the flat surface in its proximity in order to obtain quantitative information on the thermal conductivity. Measurements were averaged over 5 by 5 grid with 100 nm pitch (red spots in Figure 3) in order to decrease the experimental error due to thermal contact area fluctuations. The $T_C$-$T_{amb}$ vs $T_{NC}$-$T_C$ plots for the 6 zones (3 on the sample surface and 3 on the grains) are presented in Figure 4. We find that the thermal conductivity in zones 1, 2 and 6 approaches to that of bulk material (1.83 ± 0.29 $Wm^{-1}K^{-1}$, see Figure 1 (b)), while for the areas near the nodule (zones 3, 4 and 5), the value is about 3 to 4 times lower. Higher resolution topographic images (Figures S4-S6 in SI) demonstrate comparable RMS roughness for the area on the nodule (0.7 nm) and its proximity (0.6 nm).

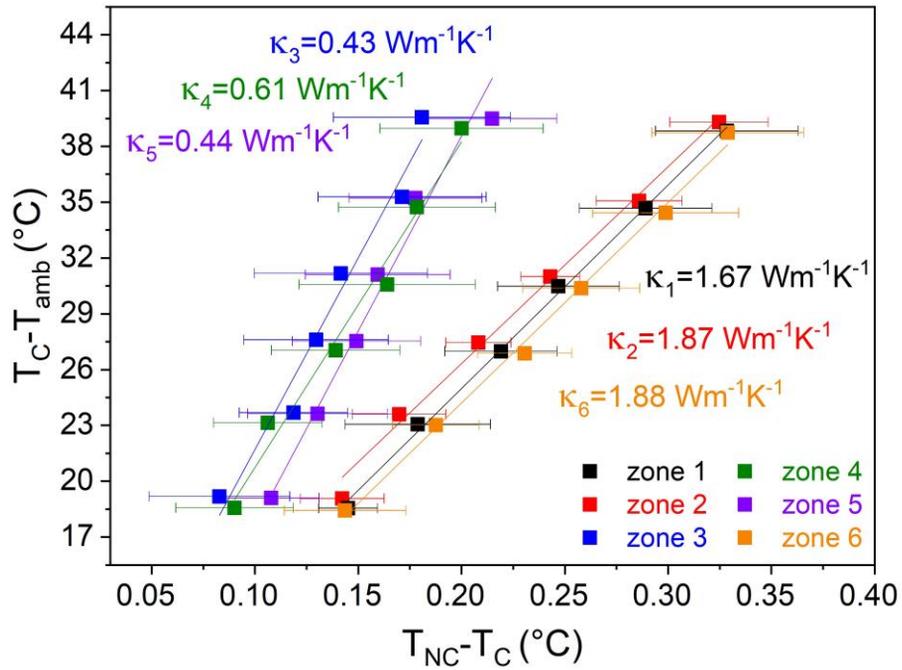

**Figure 4**: Plot of $T_C-T_{amb}$ as a function of $T_{NC}-T_C$ for the six zones presented in the Figure 3. Wheatstone bridge bias was varied between 0.7V and 1.2V in steps of 0.1V. Note that a new SThM tip was used for this series of measurements, hence the different calibration coefficients (see Figure S7 in the SI): $\alpha$=59.71 Wm$^{-1}$K$^{-1}$, $\beta$=73.98 K/K.
7

## 3.3 Comparative analysis of thermal conductivity values

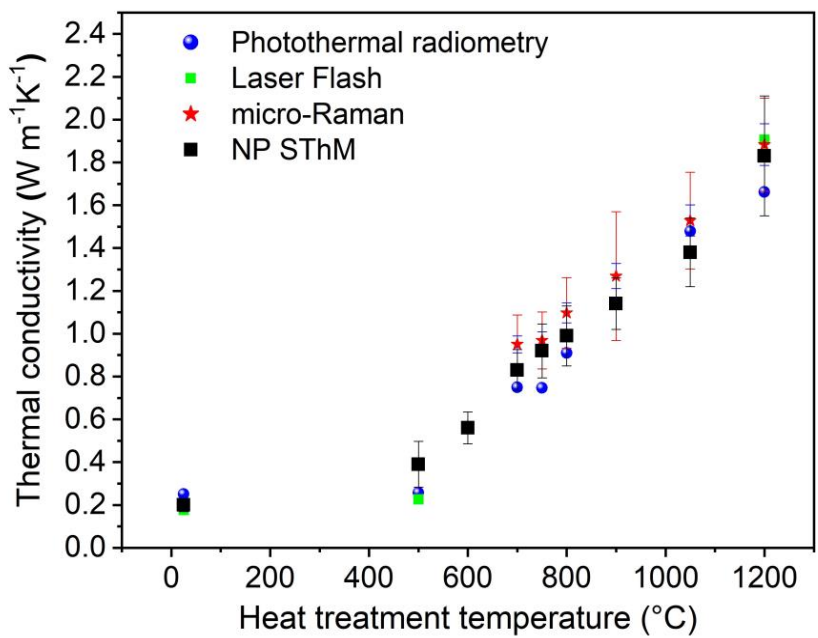

**Figure 5**: Thermal conductivity of Kapton-derived carbon as a function of heat treatment temperature. Values obtained from photothermal radiometry, laser flash analysis and micro-Raman analysis are taken from Ref.[10] for comparison.

We have compared the results of the NP SThM measurements with those obtained by other techniques[10]. For each sample, the measurement was performed in 5 different randomly selected areas with no visible topographical features on the surface in the aforementioned grid pattern. As shown in Figure 5, the SThM data show excellent agreement with the data obtained by the three other techniques (photothermal radiometry, flash laser analysis, and micro-Raman). The incertitude of the value increases with the temperature of treatment, which indicates that the thermal treatment results in a heterogeneous material, as we have previously demonstrated, which may be assigned to local variations of carbon graphitization[30] of polyimide.

## 4. Discussion

Thermal conductivity values derived from NP SThM data are validated by comparison with data obtained for bulk material by photothermal radiometry, flash laser analysis and micro-Raman thermometry. For samples heated above 900°C, a more broad distribution of values (larger error bars) is observed, consistent with previous reports in which escape of nitrogen containing species[8] from Kapton films at this temperature has been shown. These transformations lead to a spatial disorder of the chemical composition occurring on a mesoscopic scale, which makes the experimental data more distributed. If we compare it with the experimental error obtained for micro-Raman measurements[10], we may suggest that the lower resolution of this technique (laser spot diameter of 1.29 µm) does not allow discerning local distribution of thermal conductivity in contrast to SThM.

The increase of the thermal conductivity by a factor ca. 10 (upon the thermal treatment of the Kapton film) is related to the gradual increase of the concentration of $sp^2$ carbon in the film and a concomitant increase of the electrical conductivity as reported in the work of Venkatachalam *et al*[8]. Above 500°C, the electronic contribution to the thermal conductivity start dominating the phonon contribution (ca. 0.2 W m$^{-1}$ K$^{-1}$ for the as-received Kapton).

The advantage of SThM is its lateral resolution, which for micro-fabricated probes (KNT, used in this study) reaches 100 nm. It has been successfully applied by Li *et al.* to materials presenting multiple phase compositions[31,32]. To estimate the thermal conductivity, Li *et al.* have relied upon finite elements modeling to circumvent the undesirable contribution of heat flow through the surrounding air and the probe cantilever. In our case, no modeling step is required, as the calibration step accounts for both the environment (conduction through the air and the cantilever) and tip geometry (thermal contact area). Gomes *et al* have investigated the sensitivity of the SThM technique[33]. By correlating the relative Joule thermal power of the probe with the sample conductivity, they have estimated optimal sensitivity of SThM for values below tenths of Wm$^{-1}$K$^{-1}$. Therefore, Kapton-derived carbonaceous materials in this study display thermal conductivity values where the SThM sensitivity is optimal. In order to provide reliable estimation of thermal conductivity value, the sample topography must be taken into consideration[33–35]. In this regard, carbonized Kapton has low surface roughness (outside of zones with nodules) suitable for consistent thermal contact area.

Graphitization of Kapton films results in gas evacuation at temperatures higher than 600°C[8], which may be regarded as a possible reason for formation of these nodules. Temperature-induced change in Kapton films has been investigated by Mamleyev *et al*[36], where they demonstrated bubbling of film surface which is followed by release of carbon and nitrogen containing volatile species from these bubbles. Nysten *et al*[37] have demonstrated the presence of bubbles of comparable size in partially graphitized Kapton by AFM. These nodules may produce pores linked to volatile species evacuation events during heat treatment, which also results in material swelling. It is well known that grain boundaries in porous materials contribute to phonon scattering and impede thermal transport[38], which in case of nodular regions is able to overcome the expected increase of thermal conductivity due to the chemical transformation of the polymer into the $sp^2$ carbon. The lower thermal conductivity on these nodules is thus consistent with the heterogenous structure of these areas with grains and voids (air gap) as observed on SEM cross-section images (Figure S8 in the SI).

## 5. Conclusion

In summary, investigations of partially graphitized polyimide allow us to better understand the local heat-induced transformation of organic materials in inert atmosphere. In this work, we have demonstrated that the NP SThM is a powerful tool for investigating thermal properties of materials with low-to-moderate thermal conductivity. With a reliable calibration procedure, herein described, we have tested a series of samples for which the thermal conductivities span over one order of magnitude, and found that the measured values are in excellent agreement with the three other techniques: photo-thermal radiometry, flash laser analysis and micro-Raman thermometry.

We have also demonstrated that this technique is able to reveal the presence of heterogeneous surface structures in materials by probing local thermal conductivity in polyimide with incomplete graphitization degree. For material heated at 1200°C (bulk $\kappa = 1.83 \pm 0.29$ Wm$^{-1}$K$^{-1}$), we have found regions which present lower thermal conductivity ($\kappa = 0.45 \pm 0.05$ Wm$^{-1}$K$^{-1}$), which may be attributed to the presence of pores resulting from the escape of volatile species at high temperatures. The NP SThM has demonstrated its potential to differentiate surface changes with good spatial resolution.

## Supplementary material

See the supplementary material for tapping mode AFM images of the samples with derived RMS roughness, NP SThM data for calibration standards and Kapton-derived carbon samples and a SEM cross-section image of a sample treated at 1200°C.

## Acknowledgments

The authors acknowledge the financial support of the French National Research Agency (ANR), project HARVESTERS ANR-16-CE05-0029. The IEMN facilities are partly supported by Renatech. We acknowledge D. Deresmes for his valuable help with the SThM instrument.

## Conflict of interest

The authors have no conflicts to disclose.

## Data availability

The data that support the findings of this study are available from the corresponding author upon reasonable request.

# Supplementary Materials

# Nanoscale thermal conductivity of Kapton-derived carbonaceous materials

K. Kondratenko*, D. Hourlier, D. Vuillaume, S. Lenfant *

*Institut d'Electronique Microélectronique et Nanotechnologie (IEMN), CNRS, Villeneuve d'Ascq, France*

* Corresponding authors: kirill.kondratenko@iemn.fr; stephane.lenfant@iemn.fr

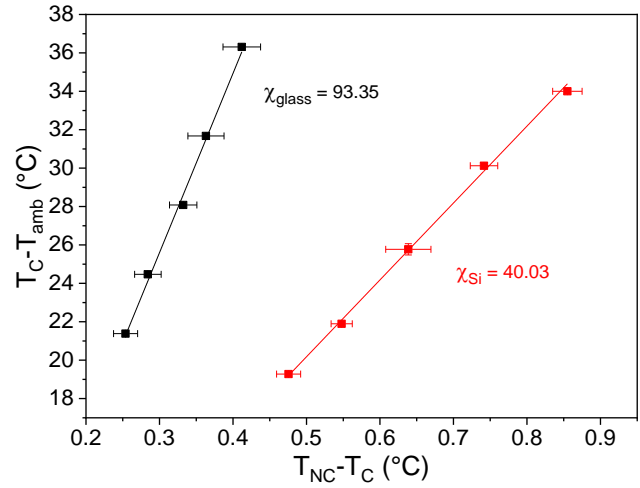

**Figure S1** Plot of $T_C$-$T_{amb}$ as a function of $T_{NC}$-$T_C$ for Si wafer and glass slide serving as calibration standard for Figure 1 (b) of the main article. χ corresponds to the slope obtained with a linear fit, from which we calculated the parameters α and β by solving the Eq. 1 (main article).

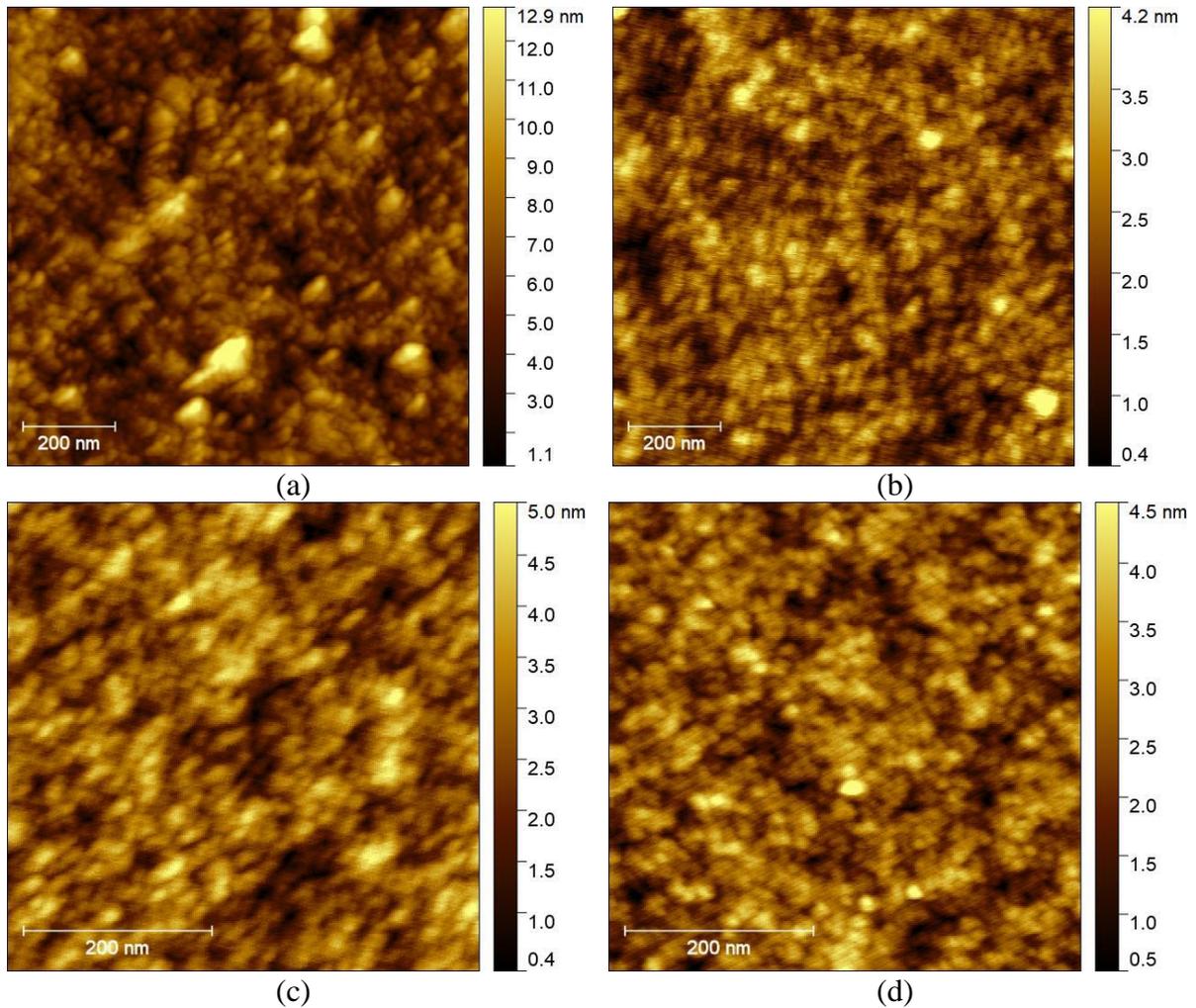

**Figure S2** AFM Topography of untreated film (a) and samples of Kapton treated at 500°C (b), 800°C (c) and 1050°C (d). RMS roughness is 1.8 nm, 0.6 nm, 0.8 nm and 0.7 nm, respectively.

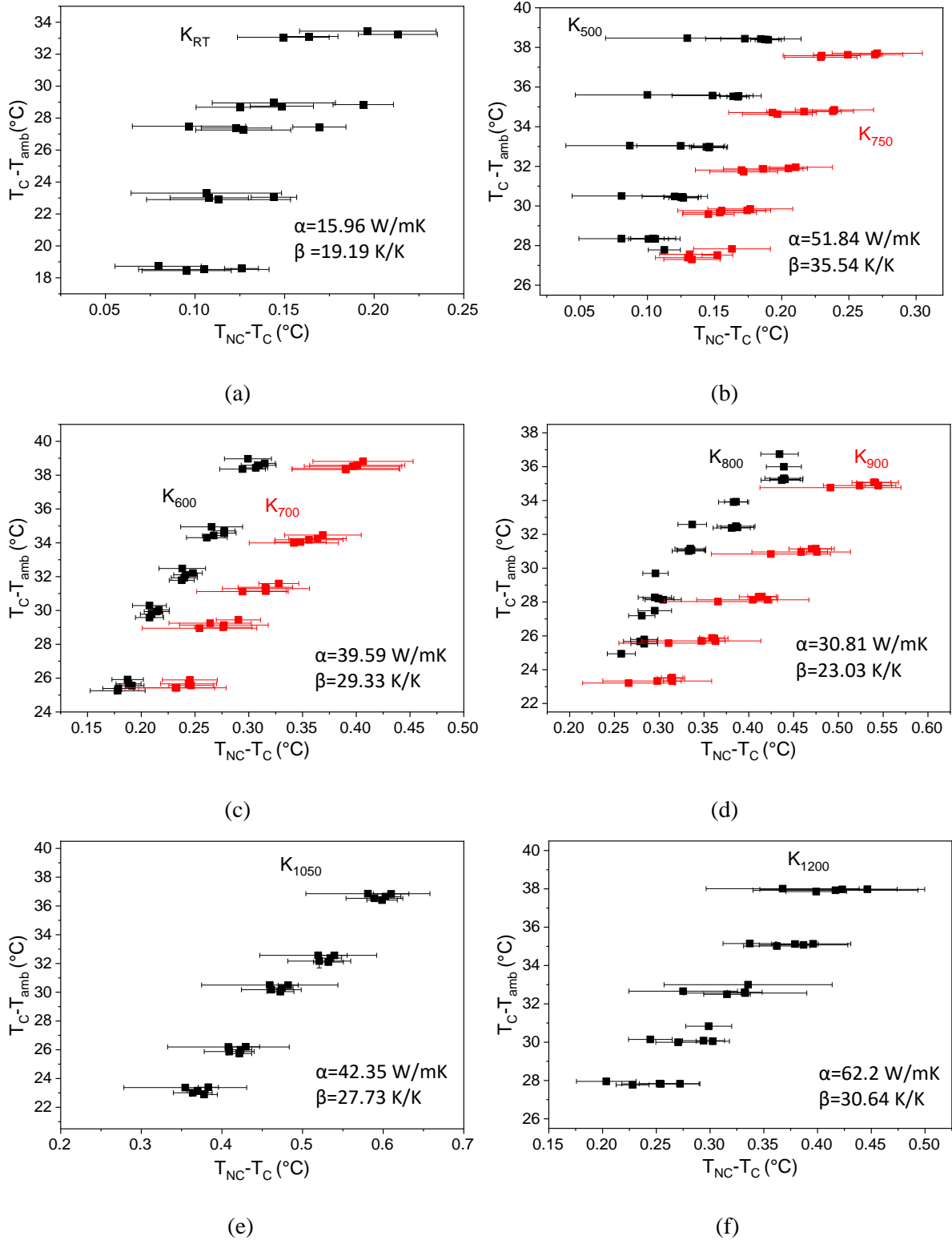

**Figure S3** NP-SThM data for untreated film (a) and samples of Kapton treated at different temperature: 500°C and 750°C (b), 600°C and 700°C (c), 800°C and 900°C (d), 1050°C (e) and 1200°C (f). Calibration coefficients are indicated in the corresponding plot. Difference between calibration parameters is due to different SThM probes. Note that additional measurements on samples treated at 500°C and 1200°C (herein presented) were performed in different measurement sessions, hence the different calibration coefficients (as compared to Figure 1(b) in the main article).

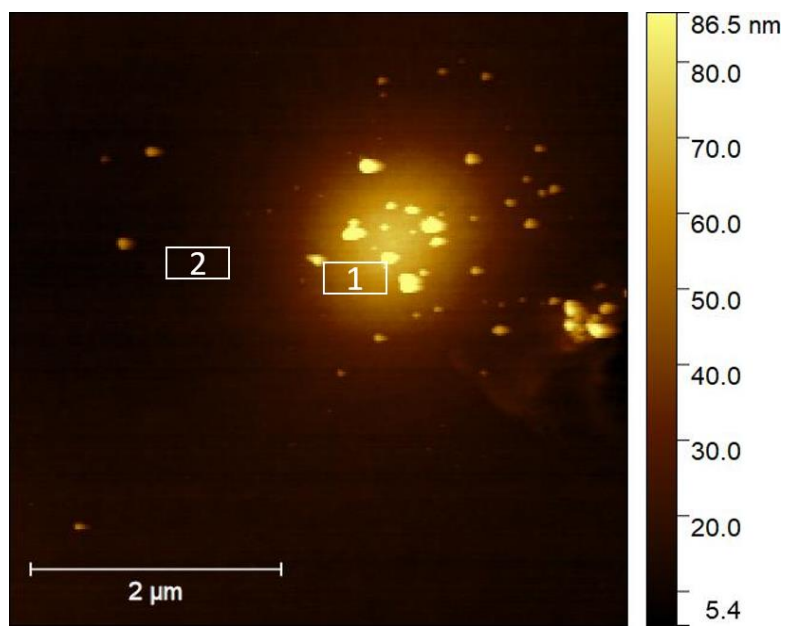

**Figure S4** Topography of a nodule (obtained by tapping mode AFM).

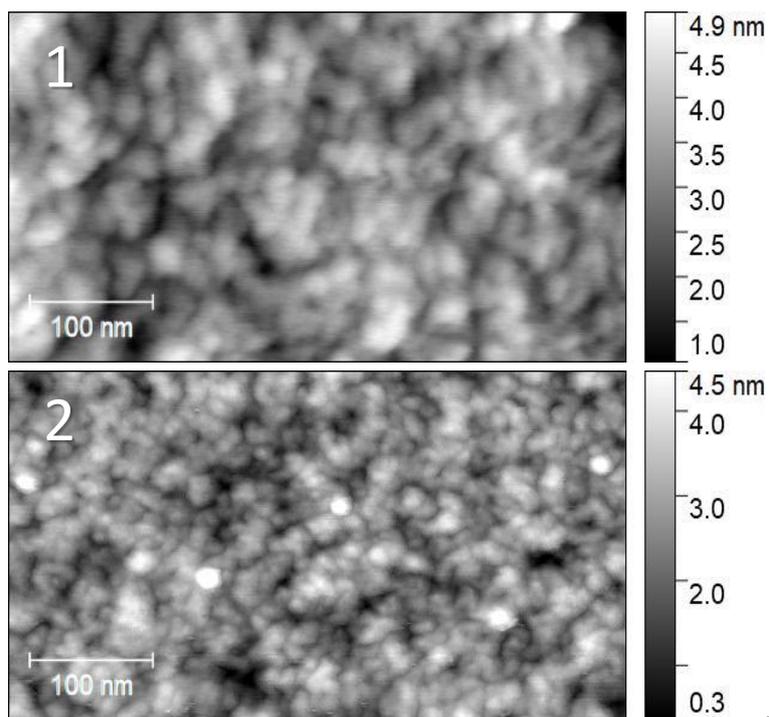

**Figure S5** High resolution topography of areas 1&2. RMS roughness values are 0.7 nm and 0.6 nm, respectively (obtained by tapping mode AFM).

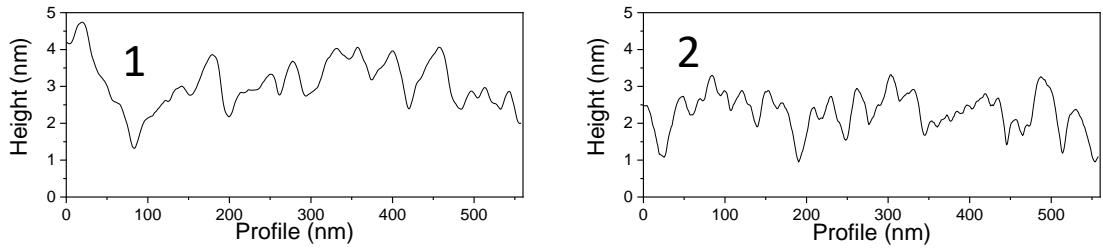

**Figure S6** Profiles extracted from Figure S4 (from upper left to bottom right corner).

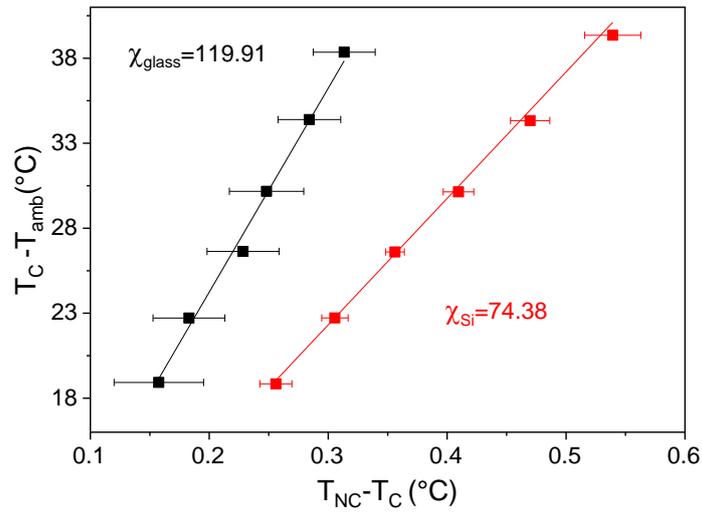

**Figure S7** Plot of $T_C$-$T_{amb}$ as a function of $T_{NC}$-$T_C$ for Si wafer and glass slide serving as calibration standard for Figure 5 of the main article. χ corresponds to the slope obtained with a linear fit, from which we calculated the parameters α and β by solving the Eq. 1 (main article).

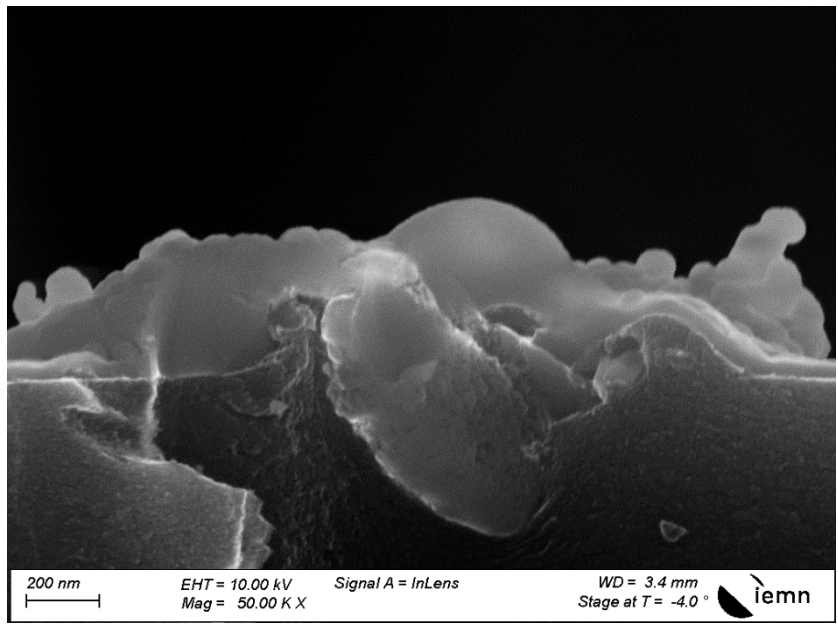

**Figure S8** SEM image of a cross section of a nodule in the sample of Kapton treated at 1200°C.